\documentclass[12pt,preprint]{aastex}

\begin{document}

\title{Faint X-ray Structure in the Crab Pulsar-Wind Nebula}

\author{F. D. Seward, W. H. Tucker}
\affil{Smithsonian Astrophysical Observatory, 60 Garden St., 
Cambridge MA 02138}

\author{R. A. Fesen}
\affil{Department of Physics and Astronomy, 6127 Wilder Laboratory,
  Dartmouth College, Hanover, NH 03755}


\begin{abstract}
We report on a Chandra observation of the Crab Nebula that gives the 
first clear view of the faint 
boundary of the Crab's X-ray-emitting Pulsar Wind Nebula, or PWN.
 There is structure in all
directions.  Fingers, loops, bays, and the South Pulsar Jet all 
indicate that either filamentary material or the 
magnetic field are controlling
the relativistic electrons.  In general, spectra soften as distance 
from the pulsar increases
but do not change rapidly along linear features.  This is 
particularly true for the Pulsar
Jet.  The termination of the Jet is abrupt; the E side 
is close to an [O {\small III}] optical 
filament which may be blocking propagation on this side.  
We argue that linear features have ordered magnetic fields
 and that the structure is
determined by the synchrotron lifetime of particles diffusing
perpendicular and parallel to the magnetic field.  We find no
significant evidence for thermal X-rays inside the filamentary envelope.

\end{abstract}


\section{The Crab Pulsar Wind Nebula}
The Crab Nebula was one of the first extrasolar sources known to emit
X-rays.  It was detected with a
rocket-borne detector on 1962 October 12 (Gursky et al. 1963),
recognized in data from a flight on 1963 April 29 (Bowyer et al. 1964a) 
and observed again
during a 1964 July 7 lunar occultation.  The occultation showed the
X-ray source to have an extent of $\sim 1^{\prime}$ 
and to be located at the center of the $5^{\prime} \times 
7^{\prime}$ optical nebula (Bowyer et al. 1964b).  The morphology of
the source has since been measured with increasingly better resolution
by a second lunar occultation (Palmieri et al. 1975; Wolff et al.
1975) and imaging
telescopes on Einstein (Harnden \& Seward 1984; Brinkmann,
Aschenbach, \& Langmeier 1985), ROSAT (Hester et
al. 1995), and Chandra (Weisskopf et al. 2000; Hester et al. 2002).
The Crab Pulsar appears as an unresolved source somewhat SE of the
bright center and the pulsar rotational energy loss 
of $4.6 \times 10^{38}$ ergs s$^{-1}$ supplies energy to the surrounding
nebula which, in turn, radiates $1.8 \times 10^{38}$ ergs s$^{-1}$ over the 
electromagnetic spectrum - from
radio to gamma rays.  The nebula's spectrum is nonthermal throughout; a photon power
law with slope of -1.25 from radio to infrared and a steeper slope of
$\approx$ -2 at the higher energies.  The power law spectrum and optical
and X-ray polarization (Hiltner 1957; Weisskopf et al. 1976) show that the 
diffuse emission surrounding the pulsar
is synchrotron radiation from highly energetic electrons moving in a 
magnetic field.  This pulsar-wind nebula, or PWN, is the brightest PWN
in our galaxy.  

The X-ray photon spectrum of the Crab PWN, $N = E^{-\alpha }$, 
in the photon energy range $E = 1-20$ keV is 
a power law characterized by $\alpha = 2.1$ 
and has been steady enough to serve as an in-flight calibration
source (Toor \& Seward 1974).  As photon
energy increases, the PWN gets smaller and $\alpha$ gradually 
increases (Strickman, Johnson, \& Kurfess 1979; Hillas et al. 1998).  
In the 1-10 kev band only 5\% of the radiation is
from the pulsar itself.  Most of the PWN X-rays come from 
a torus (Brinkmann, Aschenbach, \& Langmeier 1985) surrounding
the pulsar and is thought to be in the equatorial plane of the spinning
neutron star. A jet
aligned with the pulsar spin axis is prominent and there are
X-rays from the immediate
surroundings.  The bright torus consists of many outward moving rings
and knots (Hester et al. 1995, 2002).  There also appears to be an outward
flow but the details of energy transfer are sketchy.  Mori et al.
(2004) have derived spectra of X-rays from the torus and jet and,
because they are similar, propose that the acceleration of electrons
by the pulsar is the same in both radial and axial directions.

The transfer of energy from the pulsar to the PWN and
then to the EM spectrum was outlined by Rees \& Gunn (1974) and by
Kennel \& Coroniti (1984) using MHD models and spherical symmetry.
The relativistic wind from the pulsar is thought to expand freely until 
it terminates in a shock wave, which is necessary because the 
synchrotron nebula is confined in the cage of optical filaments 
expanding at a velocity $\sim 1500-2000$ km/s. The bright X-ray 
ring closest to 
the pulsar is presumably the location of the relativistic shock. 
At this shock the particles are
accelerated into a power-law spectrum (deduced from
the EM power-law spectrum).  
After the shock, particles flow outward radiating to form the observed
PWN and the outward flow slows to match the expansion velocity at the outer
filamentary shell.  All seaches
of a larger volume have found no direct evidence for a larger shell
(Frail et al. 1995; Seward, Gorenstein, \& Smith 2006).
  As energy is lost, the synchrotron spectrum softens and indeed  
the size of the continuum nebula increases as wavelength 
of observation increases.  

An ad hoc 2 -dimensional adaptation of the 
Kennel-Coroniti solution by Shibata et al. (2003) and two-dimension MHD 
simulations with prescribed radial energy flux density and magnetic 
field by Komissarov and Lyubarsky (2003) and Del Zanna et 
al. (2006) predict axial and equatorial 
features similar to those of the nebula. Both groups emphasize that the 
flow and magnetic field structure are complex with departures from 
spherical symmetry disrupting the classic beauty of the Rees \& Gunn and 
Kennel \& Coroniti pictures. Plasma flow is outward in the equatorial and 
polar regions but azimuthal and even inward in other sectors. Shibata et 
al. And Komissarov and Lyubarsky note that a more realistic description 
of the Crab's X-ray features will require a full 3-dimensional 
simulation that allows the development of instabilities and a disordered 
magnetic field structure.

However as discussed by Begelman (1998) these models do provide a good 
description of the Crab's gross features - a bubble inflated by a 
mixture of relativistic particles and toroidal magnetic field injected 
by the central pulsar in the form of a relativistic wind. The bubble is 
a confined cylindrical structure in which magnetic and particle pressure 
vary with cylindrical radius. Begelman further notes that this is also a 
good description of a plasma pinch with current flowing in the axial 
direction - the so-called Z-pinch - which is subject to instabilities. 
These instabilities could produce a chaotic structure with loops and 
field reversals that could lead to additional particle acceleration in 
the outer regions of the nebula, consistent with the conclusions of 
Shibata et al. (2003) and Komissarov and Lyubarsky (2003).

The present paper describes an X-ray observation of the outer edge 
of the PWN, a region
intermediate between that controlled by the pulsar and that controlled 
by the slowly expanding filamentary network.  There, X-rays are fading 
away either because of a gradual loss of energy by the radiating 
particles or by escape of the particles into a region with a lower 
magnetic field.

In these models, at any point, energy in the PWN is divided between 
magnetic field and 
relativistic particles.  The flow is
characterized by the magnetization parameter, $\sigma$ = Poynting flux
/ particle kinetic energy flux.  The magnetic field we use to calculate
particle lifetime is not this Poynting field but the equipartition
field, the field experienced by the particles without concern for
direction or time variability.

It is generally assumed that, after the shock, the energy in the PWN is
equally divided between field and particles and that the particle energy 
is equally shared
between positive and negative particles, $E_+ = E_-$.  
  The positive particles are probably positrons and radiate at
the same rate as the electrons.  If protons are present, the synchrotron
emission rate is low and, because they have not yet been observed,
their existence and fate are unknown.

The X-ray structure we report here does not convey much information about
flow in the central bright part of the nebula.  There the pulsar controls
the geometry and flow in the equatorial region where expanding rings
are observed and the initial collimation of the southern jet.  Farther
out and closer to the optical filamentary shell
 where the pulsar wind is stopped conditions
are controlled by the slowly expanding material.  Although there is
pressure from the PWN and consequent instabilities, it is the geometry
of the filaments that determines the nature of the PWN here.  However,
magnetic
fields also play a large role.  Strong nonthermal radio emission from
the Crab's optical filaments show that they have embedded fields and 
Fesen, Martin, \& Shull (1992) have argued that the synchrotron
nebula's east and west ``bays'' are caused by a toroidal field which is
a relic of the progenitor star.  We have observed the region
intermediate between that controlled by the pulsar and that controlled by
the envelope and we find that there is order in the flow.  
Conditions are not chaotic and magnetic fields seem likely to
 play an important role.  

\section{The Chandra Observation}

Because the Crab is so bright in X-rays and because the Chandra 
telemetry response
is limited, a direct ACIS observation of the Crab would result in 
 a factor of 6 deadtime.
To avoid this, we excluded the bright central region
from our observation.  The observation was
a 40 ks exposure using 3 ACIS-S  and 2 ACIS-I chips with the 
X-ray nebula centered on the S3 chip, but with the center region of
the chip excluded from the telemetry.  Thus, with dead time only a
few percent, a 40 ks exposure was
obtained of the faint outer part of the PWN. Table 1
gives the details of the observation.  This was one of two observations to
search for thermal emission and to study the Crab halo 
(Seward, Gorenstein, \& Smith 2006). \\

\begin {table}
\title{Table 1: The Chandra Observation} \\

\begin{tabular}{|l|l|l|l|} \hline\hline
Observation number& Date & Live Time& ACIS chip \\
\hline 
500432/4607 & 27 Jan 2004 & 37250 &S3 \\
\hline\hline 
\end{tabular} \\
\end{table}

This paper concerns only data taken from chip S3.
Figure 1 shows the raw image.  Note the excluded region at the center
and the decrease in intensity at the boundary of this region caused by
dither of the pointing direction.  Also note the prominent charge
transfer streak (an artifact of the ACIS detector) which makes
spectral measurements difficult east and
west of the central bright region.

Figures 2 \& 3 shows the data after subtraction of the charge transfer
streak.  In order to show the complete PWN, a short observation of the
bright Crab (obsid 1997, live time = 3972 s) has been added.  
A 2 pixel gap has been left
between the 2 observations to avoid false impressions due to time
variability or normalization problems.  The difference in surface
brightness between the brightest and faintest parts of the PWN is a
factor of $\approx 500$.

We divided the data in 2 energy bands, 0.4-2.1 keV and 2.1-8.0 keV.
Transmission through the ISM is small below 0.4 keV, there is an
abrupt change in the Chandra mirror reflectivity at 2.1 keV and
charged particle background is appreciable above 8 keV.  Structure 
at the edge of the PWN is much more prominent in the lower energy band
 because the spectra of features at the boundary are soft.  While only the 
0.4-2.1 keV data are shown in the figures, energies from 0.9 to 8.0
keV were used for the spectral analysis.

\section{Results}

\subsection{X-ray Structure}

Hester et al (1995) defined and identified
faint features at the edge of the Crab PWN as seen in a ROSAT observation.  
We see in Figure 2 all of the features 
noted by Hester et al. (in their Fig. 5 ) and more.  We have
enough counts to define structure and to extract and fit spectra.
Since we see the same features 12 years after the ROSAT observation, 
these are not transient
but time variability is certainly not excluded.  

Starting in the southeast and proceeding counterclockwise we see: the
termination of the south pulsar jet, a bright finger extending due
south,
a bright finger extending due west which is partially an extension of the
bright ring/torus structure in the central nebula, an indentation, 
the West Bay, with faint interior wisp, a faint wisp encircling the
north edge of this bay, the faint Northwest boundary (sometimes called
the "umbrella") which is perhaps
the termination of a northern pulsar jet and which appears to consist of a
number of closed loops attached to the inner nebula, a very faint
cloud north of this, a faint finger
pointed north approximately aligned with the optical [O {\small III}] 
$\lambda \lambda 4959, 5007$ ``chimney'' of filaments,
  two stubby fingers pointed northeast curving
slightly north and with striations at the base,
another indentation the East Bay, an arch of emission standing out from
the nebula which may be part of a loop structure connected to the inner
nebula, and an amorphous faint cloud of emission to the east of this
arch.

\subsection{Comparison with Optical Structure}

Figures 4 and 5 show X-ray surface brightness contours overlaid on the
optical continuum and on [O {\small III}] emission respectively.  
Contours are drawn to show particular features.
There is continuum optical emission associated with every
X-ray feature.  The Jet and the S Finger are optically bright.
There is continuum emission in the SE where the X-ray SE Cloud is
located which reinforces the nonthermal origin implied by power-law
fits to the X-ray spectra.  

Faint X-ray contours to the north roughly follow the shape of the
brightest [O {\small III]}] filaments, particularly in the Northwest.  The
Jet is bounded by [O {\small III]}] filaments on the NE side.
The S and SE Fingers extend between [O {\small III}] filaments. There is
very faint [O {\small III}] emission bounding part of the S Finger.

\subsection{X-ray Spectra}

Figures 6-10 show overlays of regions where data were extracted for spectral
analysis.  The CIAO software, version 3.2, was used for extraction and
spectral fitting and the DS9 software for imaging. 
Analysis was sometimes difficult because of pileup effects 
and because of difficulty in choosing
proper background regions. Backgrounds are not negligible.  The
brightest regions analyzed have background contributions of $\approx$
10\%.  In the faintest regions, the signal:background of
the SE Cloud was 0.7:1 and that of the N Cloud 0.2:1 (see Fig. 3).  
Parameters for spectral fits are
listed in Table 2.  Uncertainties given are formal 1$\sigma$
from the SHERPA fitting algorithm.  Systematic uncertainty arising
from choice of background region was investigated by repeating some
spectral fits using different background regions.  The resultant
change in best-fit photon index, $\alpha$, was within the formal error.

The spectrum of the Crab and the pulsar can be derived free from
pileup effects using events from the
charge-transfer (CT) streak.  Results obtained from a region just
outside the edge of the PWN are also listed in Table 2.  The photon spectral
index of $2.17 \pm 0.01$ is a bit softer than the 
photon index of $2.08 \pm 0.05$, the
standard derived from observations in the period 1964-1972 (Toor and
Seward 1974), $2.11 \pm 0.01$ measured by Willingale et al. (2001), and
$2.046 \pm 0.003$ by Kirsch et al. (2006) with XMM. 
The average column density, which was allowed to vary,
 $N_H = 3.5 \pm 0.1 \times 10^{21}$ is in good
agreement with $3.45 \pm 0.02 \times 10^{21}$  
measured by Willingale et al. but is 25\% higher than the $2.76 \pm
0.04$ measured by Kirsch et al.

Since the soft outermost part of the PWN and the 
soft dust-scattered halo are not included in the CT-streak, the
measured photon index might be expected to be harder than the 
previously-measured average but our photon index is a bit softer than average.
Our measured photon index for the Pulsar is $1.81\pm 0.05$ compared with 
$1.63 \pm 0.09$ measured by Willingale et al. and $1.72 \pm 0.05$ by 
Kirsch et al., again softer. 
We took the comparison
of the CT-streak spectra with other results as indicating that the
calibration of the ACIS instrument and CIAO software are such that
derived power-law indices should be good to within $\pm 0.1$.
For all spectral fits, $N_H$ was fixed at $3.4 \times 10^{21}$
atoms cm$^{-2}$ even though sometimes slighty better fits could be
obtained with higher $N_H$ and
somewhat higher photon spectral index.

The pileup fraction was appreciable for the brighter
regions.  A CIAO program developed by Davis (2001) was used to remove
pileup effects and we list the pileup fraction calculated by
this program for each spectrum.
Fits are good when this fraction is below $\approx$ 
15\%.  The pileup correction becomes more difficult as count rate
increases.  Because we observed that the correction failed first at the 
lowest energies, we restricted our analysis to the energy range 0.9-8.0 keV.

The selection of background regions was driven by the bright
scattering halo which extends over the entire ACIS field.  
For most regions of interest, adjacent regions are also part of
the PWN and are unsuitable for use as background.  In most cases, we
selected a region outside the PWN but at the same radial distance from
the scattering center (the center of the halo), located
14$^{\prime\prime}$
northwest  of the pulsar at R.A. = $05^h 34^m 31.3^s$, Dec. = 
$22^{\circ} 01^{\prime} 03^{\prime\prime}$ (epoch 2000).  Extraction boxes
shown in the figures were generally aligned along the circumferences
of circles centered on this point. 
Decreased exposure due to dither into the central hole in the field
was taken into account for regions at the edge of the excluded area.
We did not fit spectra to data where the exposure correction was more
than a factor of two except for the bright edge of the W Bay.
The CT streak introduces another large background in the east and
west and spectral analysis was attempted at only one place in the
area covered by the CT streak, i.e., the W Bay where a steep falloff
allowed background from the streak to be measured with
little change in radial distance.  

\subsection{Estimates of Magnetic Field Strength}

Using standard formulae (Woan 2000) for synchrotron power and characteristic
frequency, $\nu_c$, and averaging over the angle between electron
velocity vector and field line, the electron lifetime, $\tau$, is
\begin{displaymath}
\tau = 93 (h\nu_c)^{-1/2} B^{-3/2}_{-4} \; \; \mathrm{years} \hspace{.7in} (1)
\end{displaymath}
where $B_{-4}$ is in units of $10^{-4}$ Gauss
and $h\nu_c$ is in keV.

Assuming all particles
 are either electrons or positrons and that all power is radiated at $\nu_c$
 ($\approx 1$ for this observation),
the magnetic field for a given region can be calculated from the
X-ray luminosity, $L_X$, and the volume, $V$.  The parameter  
 $\beta$ = $E_{magnetic}/E_{particle}$ is the ratio of magnetic field energy 
density to particle energy density.   

\begin{displaymath}
B_{-4} = 2.5 \times 10^5 [f_p \; \beta \frac{L_X}{V}
  \frac{1}{(h\nu_c)^{1/2}}]^{2/7} \hspace{.7in} (2)
\end{displaymath}

where $L_X$ is 0.3-3.0 keV X-ray luminosity in ergs s$^{-1}$, $V$ is
volume in cm$^{3}$, $h\nu_c$ is critical energy in keV, 
and $f_p$ is a factor accounting for
particles that radiate in other wavebands.  We use $f_p$ = 400 for the
Crab which covers from radio to $\gamma$-ray 
energies.  With the usual assumption that $\beta$ = 1, ~~~
$[f_p \; \beta]^{2/7} \approx [400]^{2/7} = 5.5$.

Anticipating the discussion to follow, magnetic fields and lifetimes
of representative regions have been derived and are listed in Table 3.
Hillas et al. (1998) have calculated the field required to explain
the observed TeV spectrum.  These TeV $\gamma$-rays are the result of Compton
scattering of IR photons by the electrons that produce few-keV
synchrotron emission. They derive a central field of $1.6 \times
10^{-4}$ Gauss, a factor of 3.5 less than the average equipartition field
calculated above, and thus one might expect longer lifetimes than those
listed in Table 2.
  
\section{Discussion}

The X-ray spectrum of the entire Crab Nebula is observed to be
a power law with photon index close to
2.1 (Toor \& Seward 1974, Weisskopf et al. 2000, Willingale et al. 2001, 
Mori et al. 2004, Kirsch et al. 2006).  Furthermore, the photon index is 
$\approx$ 1.9 in the central
bright Torus and steepens to 2.3-3.0 in the outer regions.  
This is generally explained by the so-called ``synchrotron burn-off''. 
After acceleration at the shock, in the vicinity of the brightest 
inner rings, the relativistic plasma diffuses
outward radiating as it goes and forming the PWN.
The size of the PWN is observed to increase with decreasing photon 
energy as expected if the radiative lifetime of the diffusing particles 
determines the size of the nebula (see Equation 1). 

Our observation covers the edge of the synchrotron nebula where
the particle energy is dropping through the X-ray range.  With the
exception of the S Pulsar Jet, all the features to be discussed have 
 soft spectra with indices ranging from 2.8 to 4.4.
The index
of $\approx 3$ observed at the base of these features
implies a considerable
loss over particles in the bright torus.
As the outward propagation of the plasma continues, the degree of 
softening differs 
in different regions. For example, the Pulsar Jet is the most extreme case and
shows little softening along its length.  Likewise X-rays from the S and 
SW fingers have an almost constant photon spectral index along the length
of these features.  In contrast,
in the W Bay and in the N Loop regions the photon spectral index
increases rapidly as one proceeds away from the center.

\subsection{S Pulsar Jet}

In Figure 2 we made use of a previous Chandra
observation of the bright center of the Crab to show the connection of
our image with the inner part of the Pulsar jet.  
*We note that in the simulations of Komissarov and Lyubarsky (2003) and 
Del Zanna et al. (2004, 2006), the jet is formed downstream of the 
termination shock and thus does not, strictly speaking, originate from 
the pulsar. However, it is produced by the pulsar, so we will refer to 
it as the ``pulsar jet''.  The discontinuity in
brightness where the images are joined is partially from pileup 
in our observation but also probably from
time variability in brightness of the jet.  Figure 6 shows the outer 
$0.3^{\prime}$ of the pulsar jet which has broadened considerably since
leaving the vicinity of the pulsar.   
X-rays from the core of the jet are hard and there is little softening
until the S termination is reached.

 Mori et al. (2004) show spectral indices from
this region derived from a sum of the shorter Chandra exposures.
Although there are not as many events, pileup effects
are less severe in their data.  They find that the photon index is $\approx$
2.0 for the last half of the jet with spectral softening at the
sides.  They point out that soft emission from the sheath and
projection implies a core spectrum harder than measured.
In our data, the pileup correction is greatest in
the core of the jet.  Since we derive about the same photon spectral index
as Mori et al. in this area, we know that our pileup algorithm
(which is not well calibrated for a power-law spectrum)
can be used for the jet region and for regions of lesser brightness.

This feature is apparently a channel which contains the energetic
plasma from the jet.  There are distinct boundaries East, West, and
South.  At
the South edge the X-ray brightness drops a factor of 3 in a distance
of $4^{\prime\prime}$ (0.13 ly).  Although radiating strongly over 
the length of the jet, there is little change in the spectrum in the
core of the jet. 

The spectrum abruptly softens South of this
termination point implying energy loss of particles crossing the
termination.  We note that this emission may be a projection and
not associated
with the Jet in which case there is no evidence for jet particles crossing
the South boundary.

There is structure to the East side, where the spectrum softens after two
steps in decreasing surface brightness.  The Jet appears to be in a
hole in the optical nebula.
Figure 5 shows an [O {\small III}] filament
aligned with the steps at the East boundary (This is not the
brighter filament one end of which appears to just touch the S boundary).  
Lawrence et al. (1995) measure the
velocity of this feature to be +700 km s$^{-1}$ placing it in the
near half of the expanding nebula.
If interaction with this filament does indeed cause
the step structure, the S Pulsar Jet must be moving somewhat towards us.
This agrees with the finding of the theoretical models, in which the 
direction of the jet was determined from the enhanced brightness due to 
Doppler beaming.

\subsection{Other Faint Outer Features}

Other features are identified in Figures 2 \& 3 and shown in detail in
Figures 7 - 12.  The {\em S Finger} is the southernmost protrusion.  
In contrast to the jet, there is no abrupt termination and X-ray
emission fades gradually.  Particles propagate South along
this feature with little spectral change along the axis.  There is some 
softening at the edges.  One can imagine magnetic field lines stretching
NS along this feature and particles losing energy most rapidly propagating
perpendicular to the field away from the core.  Alternately, the
bright core might indicate the present position of an energy flow that
changes direction and path within the electron lifetime.  In this
case, the
softer radiation away from the core would be from particles deposited
previously.  Figure 5 shows very faint [O {\small III}] wisps bounding
the sides of this feature, implying an interaction with the envelope in
this area.  This is the Hester et al. (1995) feature `OS'.

Directly West of this the {\em SW finger} starts following the direction of 
the inner loops but keeps
heading west instead of turning north to circle the pulsar.  The 
spectrum, appreciably softer than that of the Torus, changes 
little along the bright length of this feature
but softens slightly in the transverse direction and at the very end.
The bright core of this feature ends in a little loop.  There is faint
emission West and South of this.  This is the Hester et al. (1995) 
feature 'OSW'.
The striations at the base of this feature (Fig. 8) follow the rings
of the bright torus (Fig. 2), as if the finger were drawn out of the
torus structure.

The X-ray {\em E and W Bays}
correspond to the well-known optical synchrotron bays 
(Fesen, Martin \& Shull 1992).
Spectral analysis is difficult because both are overlaid by the
CT streak.  We did make spectral fits to 2 regions in the
W Bay as shown in Figure 9.  The W Bay is formed by a sudden decrease
(factor of 2) in emission as one proceeds outward.  A fainter feature
lies projected within this bay and this too ends with a sudden decrease (factor of 4) in
emission.  The softer spectrum of this feature shows that it is
indeed a separate region and not a continuation of the first step
seen in projection.

The {\em W Finger}
is faint and wraps around the North side of the W Bay.  It
also seems to merge with faint structure at the northwest corner of the
nebula.

The {\em N Cloud} is just North of the NW Loops and 
barely visible above background.  There are too few
events to define any structure but a rough spectral analysis was possible
using a signal of 2000 events over 10,000 background events.  
Best fits are a power law with photon index $\approx 4.4$ or a thermal
bremsstrahlung continuum with $kT \approx 0.35$ keV.  There is no
significant spectral structure indicating emission lines. 
Of all the X-ray features described here, it is the closest to the
optical boundary and can be seen in Figure 3 just inside the [O {\small III}]
edge of the nebula. In this area, Fesen, Martin, \& Shull (1992) noted faint
optical continuum emission outside the filamentary envelope.  This can
be seen in their Figure 8.  It does not overlap with the N Cloud.

The {\em NW Loops} form the northwest edge of the nebula.
This region is the area called the 'umbrella' by Mori et
al. (2004), and the 'counterjet' 
by Hester et al (1995).  The region
is filled with faint striations or loops oriented roughly parallel to
the circumference of the PWN.  Our spectral extraction regions are
oriented along this structure.  The spectrum softens appreciably
going outwards, indicating rapid energy loss.
This is Hester et al. (1995) feature 'ON'.

The {\em Chimney} is a very
 faint finger pointed north [Hester et al. (1995) feature 'ONE']
and approximately at the base of the faint 
north-pointing optical [O {\small III}] ``chimney'' (van den Bergh
 1970, Gull \& Fesen 1982, Fesen \& Gull 1986).  The
alignment of the axis is somewhat east of the optical chimney axis and the
X-ray emission fades before reaching the base of the optical chimney.  
Contrary to the result for the S Finger, there
is some spectral softening as one proceeds north, but uncertainty in
 the outermost spectral index is large.  It is possible to imagine
particles streaming north along this feature and then along the
optical chimney but the connection between this feature and chimney 
is not seen in the X-rays, with weak evidence for an optical
 synchrotron connection as well (Woltjer \& Veron-Cetty 1987). 

The {\em NE Finger}
has 2 striations across the base, each an $\sim$ 30\%
increase in surface brightness over adjacent emission.  The feature
then splits into 2 parts and fades.  This could be 2 separate
features seen in projection.  Because this is faint and embedded in the CT
streak, derived spectral parameters had large uncertainties and
are not listed.  Looking at Figure 2, one can imagine this feature to
be a warped extension of the bright torus around the pulsar.  The
striations line up with the rings that have been observed to flow
outward.  Perhaps these are an extension of the ring phenomenon. 

The {\em SE Arch}
only appears below 2 keV in energy.  It has three
components which may or may not be connected.  The northern side of
the arch is a wisp which borders the S side of the E Bay.  The
structure is similar to that of
the W Finger which wraps around the N side of the W Bay.  The S side
of the arch is possibly the E end of a feature which passes south 
of the Pulsar Jet
termination and joins the S Finger.  The E side of the arch consists
of 2 elongated knots which may be associated with separate loops
connecting to the inner nebula.

The {\em SE Cloud} is somewhat separated from the main PWN. 
The structure here is less ordered than that of other features. 
Although faint, it is clearly seen and a good spectrum was extracted. 
The spectrum follows the general rule that features farthest from the 
center have the softest spectra.
The amorphous structure at first
looked promising as a sought-after region of thermal 
emission but we found no
lines in its spectrum which would identify the
emission as thermal.  A power law fit is good and the presence of
optical continuum in this area leads to the conclusion that this is
also synchrotron emission.  This is Hester et al. (1995) feature OSE.

\subsection{General}
  
The Crab Nebula's PWN is
confined by the remnant's optical
filamentary envelope and the X-ray synchrotron features discussed here
showcase places where the confining pressure is lowest.  The E and W
bays seem to wrap around massive optical filaments.  The S Pulsar 
Jet seems formed by
an [O {\small III}] filament on the E side.  The S and SW Fingers 
are located in
spaces between bright filaments.  The northern edge of the PWN falls
just inside the bright northern filaments and tracks their morphology.

We propose that the features such as jet, loops, fingers, are
 defined by or at least contain ordered magnetic fields.  
Electrons flow along the features more rapidly than
perpendicular to the field lines.  

We assume that the width of the filaments is determined by the 
distance an electron can diffuse perpendicular to the magnetic field in 
a synchrotron lifetime, $\tau$ (cf. The discussion of nonthermal filaments in 
supernova shock waves by Vink (04)). Using standard diffusion theory, 
the width, $W$, is given by

$W = (6 \; D \; \tau )^{1/2}$ \hspace{4.0in}(3)

\noindent where $D$ is the diffusion coefficient for diffusion perpendicular to the 
magnetic field. The general assumption (e.g., Vink 2004, Yamazaki et 
al. 2004, Reynolds 1998, and Jokipi 1987) is that

$D = R_g \; c \; /3\eta$

\noindent where $1/\eta$ is the mean free path in terms of gyro
radii, $R_g$, and $1< \eta < 10$.  Then

$W = (2 \; c \; R_g \; \tau \; /\eta )^{1/2} = 2.3 \times 10^{15} \;
\eta ^{-1/2} \;  B_{-4}^{-3/2}$ \hspace{2.0in}(4)

= $7.7^{\prime\prime} \; \eta ^{-1/2}  \; B_{-4}^{-3/2}$ at 2 kpc.

\noindent For $B = 5 \times 10^{-4}$ and $\eta = 3, W = 0.4^{\prime\prime}$, 
comparable to the observed fall-off distances.

The Chandra image data reveal rapid decreases in surface brightness 
at the end of the S
Pulsar Jet, in the W Bay and along the N edge of the SW finger.
In these regions, brightness drops a factor of $\approx 3$ in a distance of
 $\approx 2^{\prime\prime}$.  
This is likely an indication of the reduced diffusion length in a 
synchrotron lifetime for electrons diffusing perpendicular to the 
magnetic field.

Although the S and SW Fingers are probably shaped by the structure of 
the optically bright filamentary
material, bulk plasma flow along these features with velocity c/2 would
take $\sim 3$ years to traverse the length of the features,
comparable to the particle lifetimes.  Yet, the photon spectral index remains
constant within the uncertainties listed in Table 2.  The electrons
have certainly lost energy coming from the inner nebula to the base of
these fingers but there is little loss along the length.  In the North and
West, the flow is across field lines and, because the spectral index 
increases, the time required to do so must be a large fraction of the 
lifetime.  Indeed, particle lifetimes may determine the extent of these
features. 

\section{Summary}

Using the Chandra X-ray image, we have examined a variety of 
features at the edge of the Crab Nebula's PWN.
We find no indication of thermal radiation and the power-law spectra
indicate that 'synchrotron burnoff' is the main mechanism for energy loss. 
This, of course, is well known as is the fact that spectra generally 
get softer as energy flows outward in the nebula.  However, we find that this
is also true on a fine scale and that losses are less along linear features.
There is also an order in the flow 
between the inner pulsar-dominated and the outer filament-dominated 
regions.  This flow is not chaotic and may be magnetically controlled. 
The dimensions of some outlying features can be understood in 
terms of particle lifetimes.
Magnetic fields anchored in the
filamentary envelope may play an important role in guiding the
X-ray-emitting particles, starting at about half the distance from the
pulsar to the outer boundary.   

This work was supported by Chandra Grant GO4-5059X.
\section{References}

Begelman, M.C. 1998, ApJ, 493, 291

Bowyer, S., Byram, E., Chubb, T., \& Friedman, H. 1964a, Nature, 201, 1307

Bowyer, S., Byram, E., Chubb, T., \& Friedman, H. 1964b, Science, 146, 912

Brinkmann, W., Aschenbach, B., \& Langmeier, A. 1985, Nature, 313 662

Davis, J.E. 2001, ApJ, 562, 575

Del Zanna, L. , Amato, E., Bucciantini, N. 2004, A\&A, 421, 1063

Del Zanna, L. , Volpi, D., Amato, E., Bucciantini, N. 2006, A\&A, 453, 621

Fesen, R.A. \& Gull, T.R. 1986, ApJ, 306, 259

Fesen, R.A., Martin, C.L. \& Shull, J.M. 1992, ApJ, 399, 599

Frail, D.A., Kassim, N.E., Cornwell, T.J., \& Goss, W.M. 1995, ApJ, 454 L129

Gull, T.R. \& Fesen, R.A. 1982, ApJ, 260, 75

Gursky, H., Giacconi, R. \& Paolini, F. 1963, Phys Rev Letters, 11, 530

Harden, F.R. \& Seward, F.D. 1984, ApJ 283, 274

Hester, J.J., et al. 1995, ApJ, 448, 240

Hester, J.J., Mori, K., Burrows, D., Gallagher, J.S., Graham, J.R.,
Halverson, M., Kader, A., Michel, F.C. \& Scowen, P. 2002, ApJ Letters,
577, L49
,
Hillas, A.M., et al. 1998, ApJ, 503,744

Hiltner, W.A. 1957, ApJ, 125, 300

Jokipii, J. 1987, ApJ 313, 842

Kennel, C.F. \& Coroniti, F.V. 1984, ApJ 283, 710

Kirsch, M.G.F., Schonherr, G., Kendziorra, E., Freyberg, M.J., Martin,
M., Wilms, J., Mukerjee, K., Breitfellner, M.G., Smith, M.J.S. \&
Staubert, R. 2006, A\&A, to be published (astro-ph/0604097)

Komissarov, S.S., \& Lyubarsky, Y.E. 2003, MNRAS, 344, L93

Lawrence, S.S., MacAlpine, G.M., Uomoto, A., Woodgate, B.E., Brown,
L.W., Oliversen, R.J., Lowenthal, J.D. and Liu, C. 1995, AJ, 109, 2635 

Mori, M., Burrows, D.N., Hester, J.J., Pavlov, G.G., Shibata, S., \&
Tsunemi, H. 2004, ApJ, 609, 186

Palmieri, T.M., Seward, F.D., Toor, A., \& Van Flandern, T.C. 1975,
ApJ, 202, 494

Rees, M.J. \& Gunn, J.E. 1974, MNRAS 167, 1

Reynolds, S. 1998, ApJ 493, 375

Schaller, E. L., \& Fesen, R. A., 2002, AJ, 123, 941

Seward, F.D., Gorenstein, P., \& Smith, R.K. 2006, ApJ, 636, 873

Shibata, S., Tomatsuri, H., Shimanuki, M., Saito, K. \& Mori, K. 2003,
MNRAS, 346, 841

Strickman, M.S., Johnson, W.N., \& Kurfess, J.D. 1979, ApJ Letters, 230, L15

Toor, A. \& Seward, F.D. 1974, AJ, 79, 995

van den Berg, S. 1970, ApJ, Letters 170, L27

Vink, J. 2005, in High Energy Gamma-Ray Astronomy: 2nd International 
Symposium, Edited by Felix A. Aharonian, Heinz J. Völk, and Dieter 
Horns. 160-171 (astro-ph 0409517)

Weisskopf, M., Cohen, G., Kestenbaum, H., Long, K., Novick, R. \&


Weisskopf, M., et al. 2000, ApJ Letters, 536, L81

Willingale, R., Aschenbach, B., Griffiths, R.G., Sembay, S., Warwick,
R.S., Becker, W., Abbey, A.F., \& Bonnet-Bidaud, J.-M. 2001, A\&A, 365, L212

Woan, G., 2000, {\it The Cambridge Handbook of Physics Formulas}
Cambridge University Press, Cambridge

Wolff, R.H., Kestenbaum, H.L., Ku, W., \& Novick, R. 1975, ApJ, 202, L15

Woltjer, L., \& Veron-Cetty, M.P. 1987, A\&A, 172, 7

Yamazaki, R. et al. 2004, A\&A. 416, 595

\newpage
\begin{table}
\scriptsize
\title{Table 2: Power-law Fits}\\
\begin{tabular}{|l|l|ll||l|l|ll|} \hline\hline
region & photon & pileup & goodness& region & photon & pileup & goodness \\
       & index  & fraction & of fit &  & index  & fraction & of fit \\ \hline 
jet 1 & $3.05 \pm .08$ &0.02&1.14&swfing 1 & $2.97 \pm .04$ &0.18&2.08\\ 
jet 2 & $2.77 \pm .07$ &0.09&1.01&swfing 2 & $3.04 \pm .04$ &0.09&1.22\\ 
jet 3 & $2.73 \pm .06$ &0.12&0.85&swfing 3 & $2.85 \pm .04$ &0.12&1.95\\
jet 4 & $2.72 \pm .05$ &0.12&1.26&swfing 4 & $3.07 \pm .03$ &0.08&1.05\\
jet 5 & $2.48 \pm .05$ &0.16&1.03&swfing 5 & $2.84 \pm .04$ &0.11&1.91\\
jet 6 & $2.30 \pm .03$ &0.17&1.42&swfing 6 & $3.10 \pm .03$ &0.05&1.46\\
jet 7 & $2.14 \pm .04$ &0.21&2.22&swfing 7 & $2.84 \pm .03$ &0.07&1.28\\
jet 8 & $2.34 \pm .04$ &0.17&1.88&swfing 8 & $2.84 \pm .04$ &0.02&0.67\\
jet 9 & $2.23 \pm .02$ &0.17&1.47&swfing 9 & $3.05 \pm .02$ &0.00&0.88\\
jet 10& $1.97 \pm .02$ &0.20&2.58&swfing 10& $2.99 \pm .06$ &0.02&1.11\\
jet 11& $2.24 \pm .05$ &0.17&1.75&         &                &    &    \\
jet 12& $2.36 \pm .02$ &0.13&1.71&sfing 1 & $2.96 \pm .04$ &0.07&1.21\\
jet 13& $2.19 \pm .02$ &0.15&1.96&sfing 2 & $2.96 \pm .02$ &0.10&1.23\\
jet 14& $2.37 \pm .05$ &0.11&1.37&sfing 3 & $2.96 \pm .03$ &0.07&1.09\\
jet 15& $3.12 \pm .07$ &0.03&1.00&sfing 4 & $3.24 \pm .04$ &0.03&0.91\\
jet 16& $3.32 \pm .13$ &0.03&0.75&sfing 5 & $3.01 \pm .03$ &0.05&0.99\\
jet 17& $3.41 \pm .08$ &0.01&0.94&sfing 6 & $2.98 \pm .05$ &0.03&0.93\\
jet 18& $3.28 \pm .14$ &0.00&0.93&sfing 7 & $2.98 \pm .03$ &0.03&1.33\\
      &                &    &    &sfing 8 & $2.98 \pm .05$ &0.03&1.03\\
chim 1 & $3.32 \pm .06$ &0.01&0.94&sfing 9 & $3.43 \pm .07$ &0.02&0.79\\
chim 2 & $3.67 \pm .08$ &0.01&0.89&&&&\\
chim 3 & $3.96 \pm .20$ &0.00&1.07&wbay 1  & $2.08 \pm .02$ &0.11&1.86\\
        &               &    &    &wbay 2  & $3.21 \pm .04$ &0.03&1.60\\
nwloop 1& $3.40 \pm .03$ &0.07&1.19&&&&\\
nwloop 2& $3.60 \pm .03$ &0.02&0.93&secloud & $3.89 \pm .08$ &0.00&0.96\\
nwloop 3& $3.80 \pm .03$ &0.00&1.52&search 1& $3.62 \pm .17$ &0.00&0.65\\
nwloop 4& $4.13 \pm .07$ &0.03&1.58&search 2& $3.76 \pm .08$ &0.00&0.92\\
nwloop 5& $3.37 \pm .02$ &0.03&1.58&&&&\\
&&&                                &ave crab& $2.17 \pm .02$ &    &0.96\\
ncloud  & $4.45 \pm .46$ &0.00&0.87&pulsar  & $1.81 \pm .05$&&1.16\\
\hline\hline 
\end{tabular} 
\end{table}
\normalsize
\begin{table}
\title{Table 3: Magnetic Fields and Lifetimes}\\
\begin{tabular}{|l|l|l|l|} \hline\hline
region&field&lifetime&extent of region \\ \hline
& ($10^{-4}$ Gauss) & (years) & (light years) \\ \hline
PWN average &5.8&6.7 &1.9 ($60^{\prime\prime}$ radius) \\
bright Torus&7.7&4.3& 1.7 (radius)\\
NW Loops &9.1&2.9& 0.6\\
Jet center, region 7&9.1&2.9&1.8 (length of jet) \\
bright inner ring&10.8&2.6& 0.08 (thickness of ring)\\
knot in bright inner ring&15.3&1.5& 0.07 (size of knot)\\
S Finger, region 5&6.2&5.9& 1.3 (length of finger)\\
SW Finger, region 7&6.2&5.9& 1.6 (length of finger)\\ \hline
\end{tabular}
\end{table}

\newpage
\begin{figure}
\center
\includegraphics[totalheight=3in]{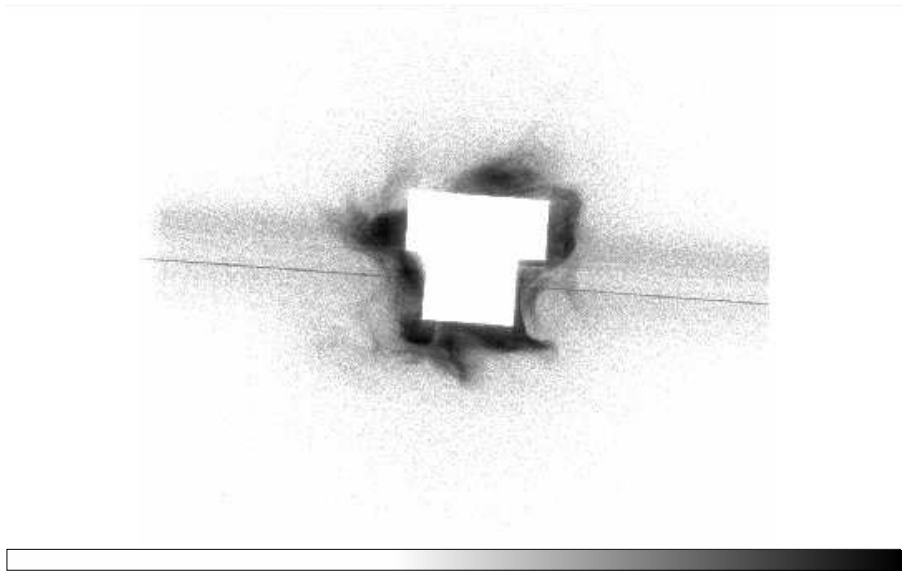}
\caption{Raw Chandra image of the Crab.  The bright center has been
 excluded from the
  telemetry but the charge-transfer (CT) streak from this region extends
  prominently east-west. Extent of the streak in this figure is
  8.0$^{\prime}$. }
\end{figure}

\begin{figure}
\center
\includegraphics[totalheight=3in]{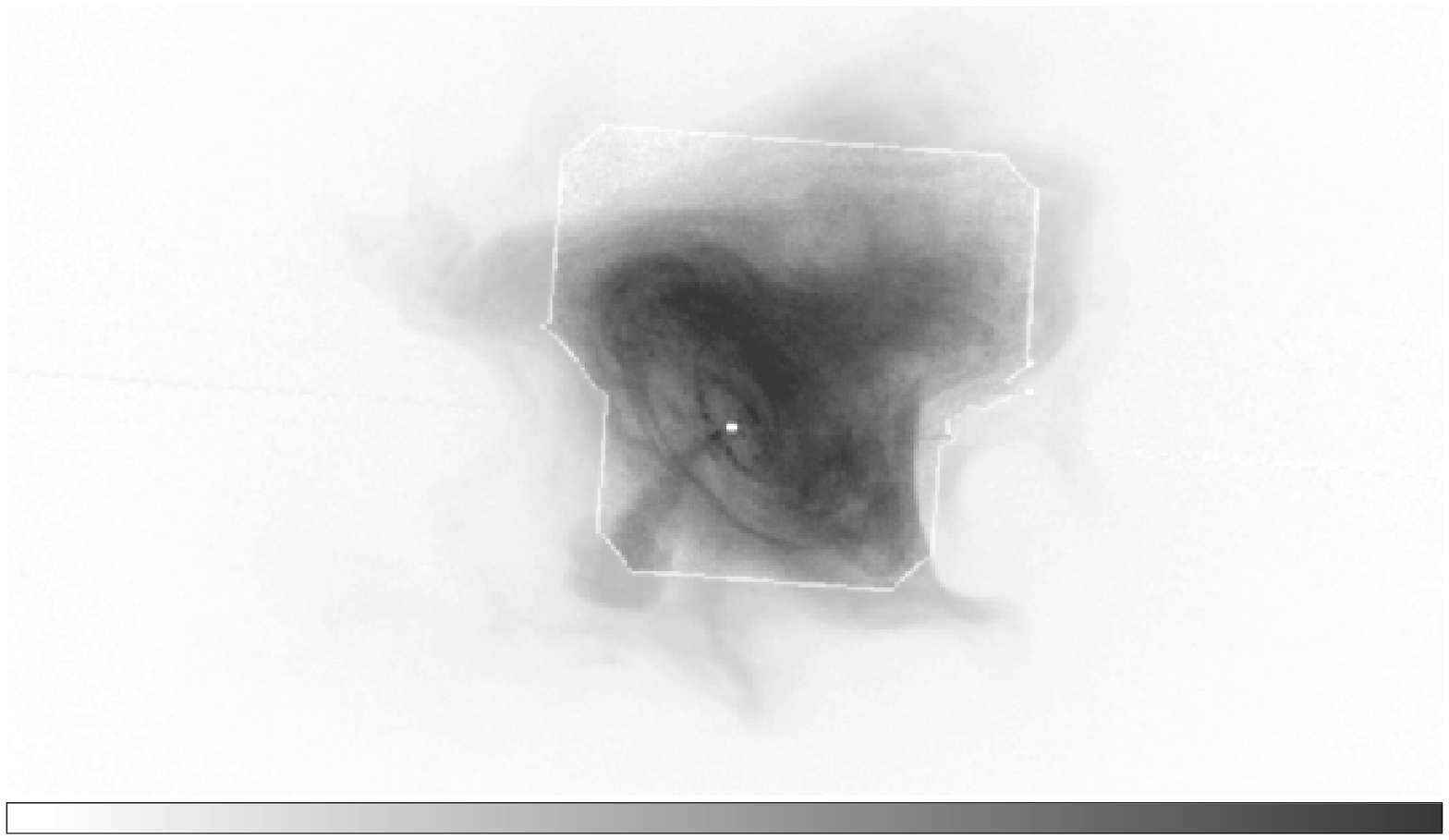}
\includegraphics[totalheight=3in]{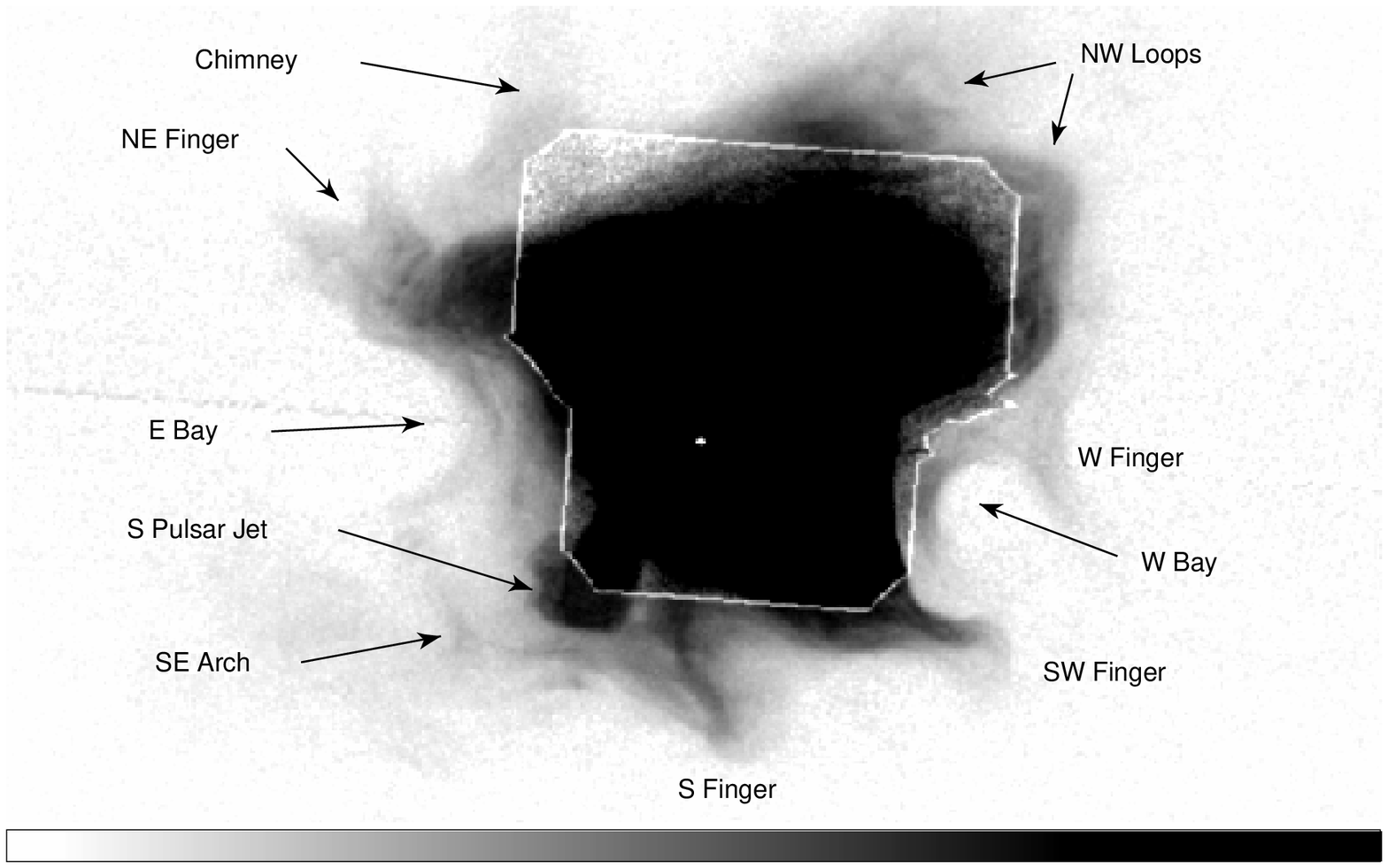}
\caption{(a)The Crab PWN in the energy range 0.4-2.1 keV.  The gap
separates this observation from an archival observation of the bright
center which has been added. Data have been smoothed using a Gaussian with 
fwhm = $0.8^{\prime\prime}$.
(b) The same image with center overexposed to show fainter features.
Each picture has a vertical extent of 3.6$^{\prime}$}
\end{figure}

\begin{figure}
\center
\includegraphics[totalheight=3in]{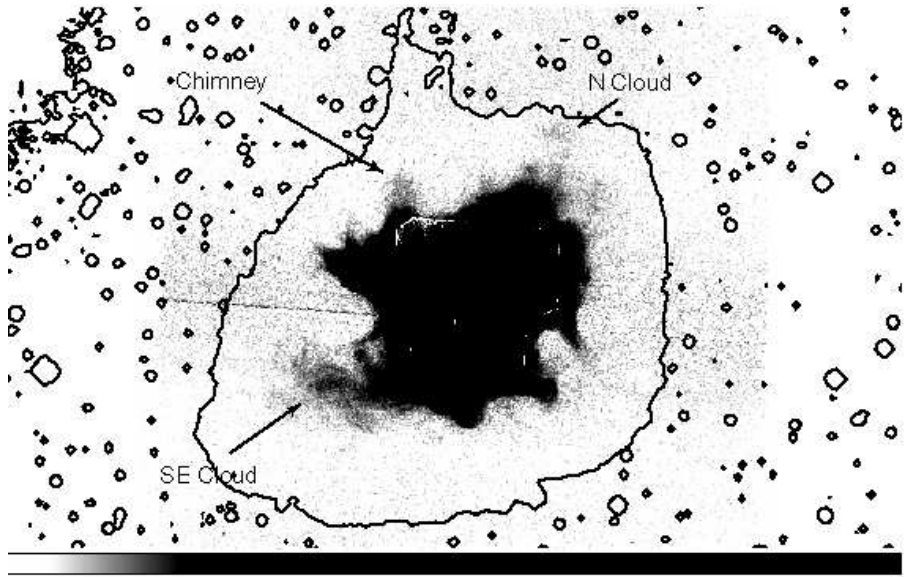}
\caption{The Crab PWN in the energy range 0.4-2.1 keV 
displayed to show the faintest features.  Those which are
hard to see in the previous figure are identified here.  A contour showing 
maximum extent of [O {\small III}] emission from the Nebula is
superposed.  The vertical extent of this figure is 7.3$^{\prime}$.}
\end{figure}

\begin{figure}
\center
\includegraphics[totalheight=3in]{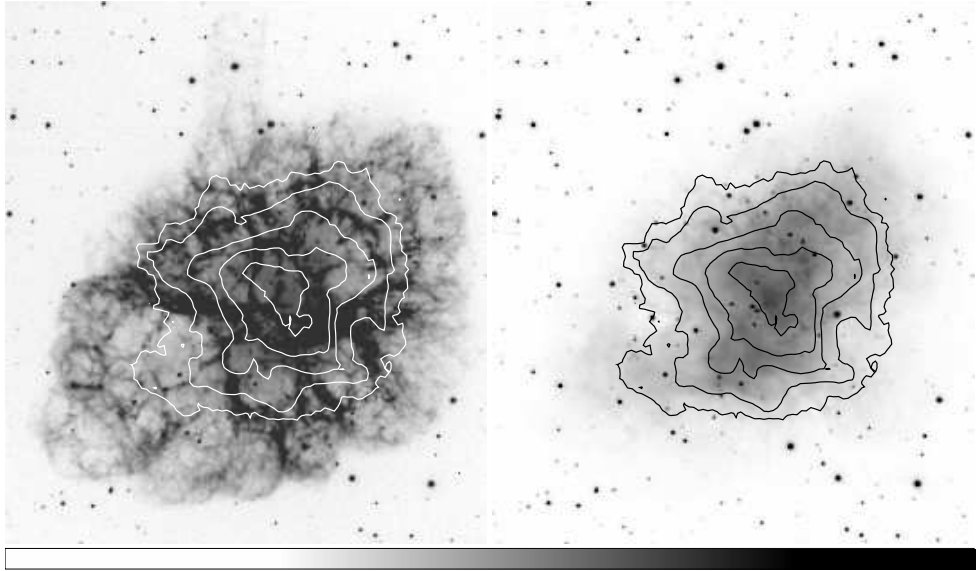}
\caption{Contours of constant X-ray surface brightness overlaid on
the [O {\small III}] filamentary emission and on the $6450$ \AA ~ optical 
continuum.  Contours are spaced to show the very
faint X-ray emission, the termination of the S Pulsar Jet, and the bright
central region.  Contours are at 0.4, 0.9, 7, 31, and 74\% of the 
brightest diffuse X-ray emission.  The vertical extent of this figure
is $7.3^{\prime}$}
\end{figure}

\begin{figure}
\center
\includegraphics[totalheight=3in]{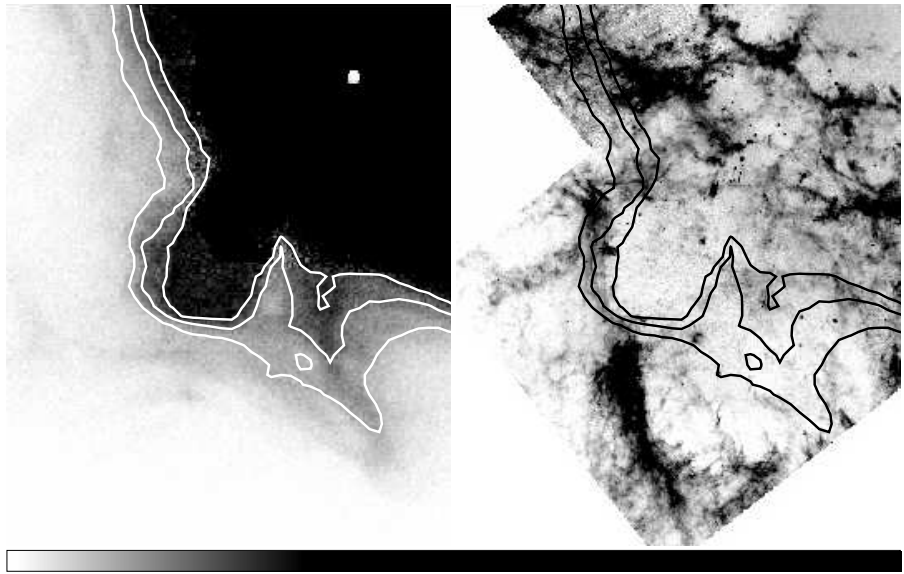}
\caption{Detail of S Pulsar Jet termination. Each picture is
  $1.6^{\prime} \times 1.4^{\prime}$ on a side. Contours of constant 
X-ray surface brightness are overlaid on X-ray (left) and [O {\small III}]
(right) images.  Contours are at 4, 7, and 10\% of the 
brightest diffuse X-ray emission.  A faint [O {\small III}] filament
  follows the NE boundary of the X-ray jet.  HST image from
  Schaller \& Fesen (2002)}
\end{figure}

\begin{figure}
\center
\includegraphics[totalheight=3in]{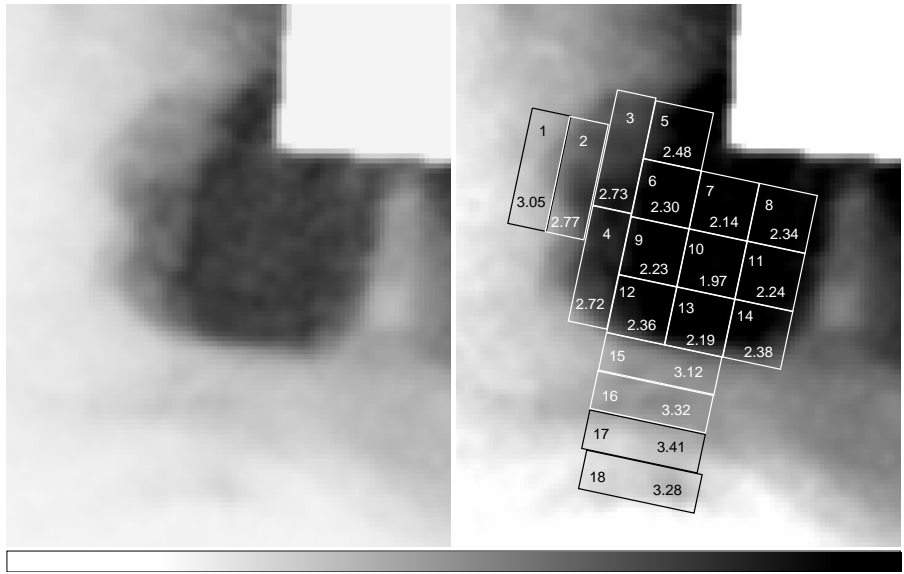}
\caption{The termination of the S Pulsar Jet showing structure and
regions used for spectral extraction. The square regions are
$5.9^{\prime\prime}$ or 10 light-weeks on a side.
The left image is displayed with a linear
stretch and the right with a square-root stretch.  Integers indicate the
region identification listed in Table 2.  Decimals are values of the
power-law photon index for the region.  This image (used in Figures 6-10) is
the raw data of Figure
1 divided by an exposure map and smoothed.  The upper right corner has
been excluded from the telemetry.  Because of dither, 
exposure is low close to this central blackout area.} 
\end{figure}

\begin{figure}
\center
\includegraphics[totalheight=3in]{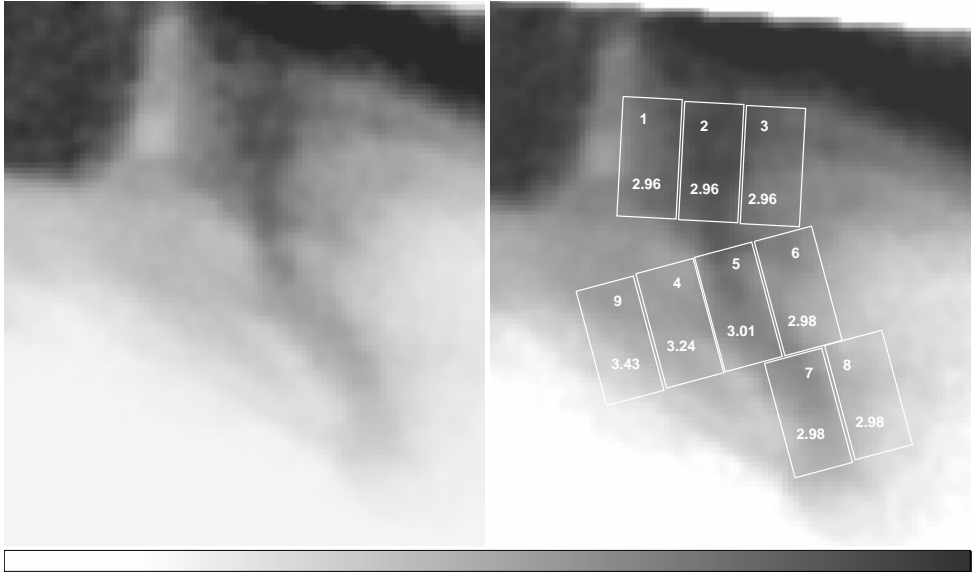}
\caption{The S Finger showing structure and
regions used 
for spectral extraction. Regions are $5.9^{\prime\prime}
\times 11.8^{\prime\prime}$ 
or $10 \times 20$ light-weeks in size.  See caption of Figure 6 for
other detail.  The photon index is approximately constant along the
length of the core of this feature}
\end{figure}

\begin{figure}
\center
\includegraphics[totalheight=2in]{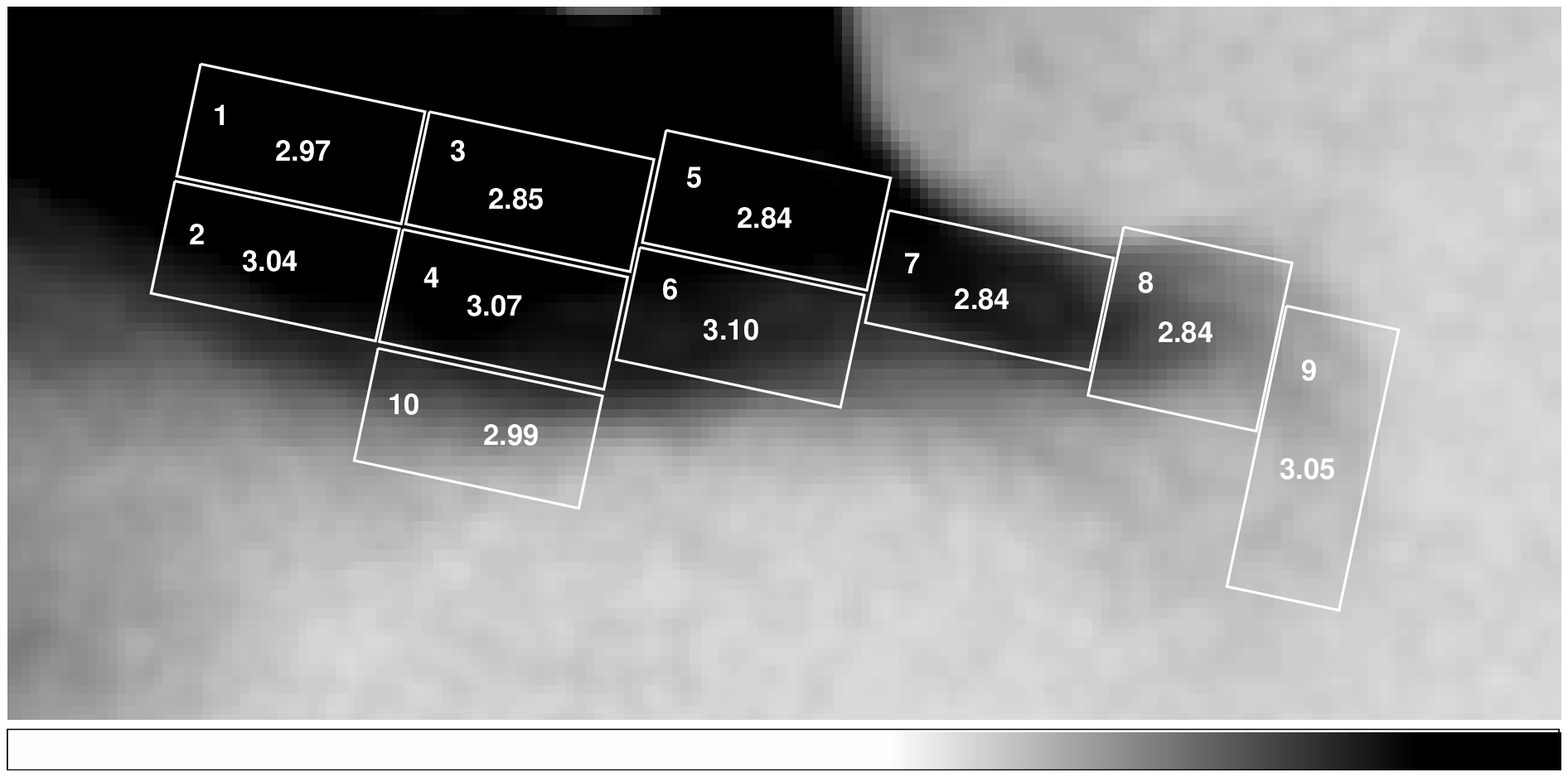}
\includegraphics[totalheight=2in]{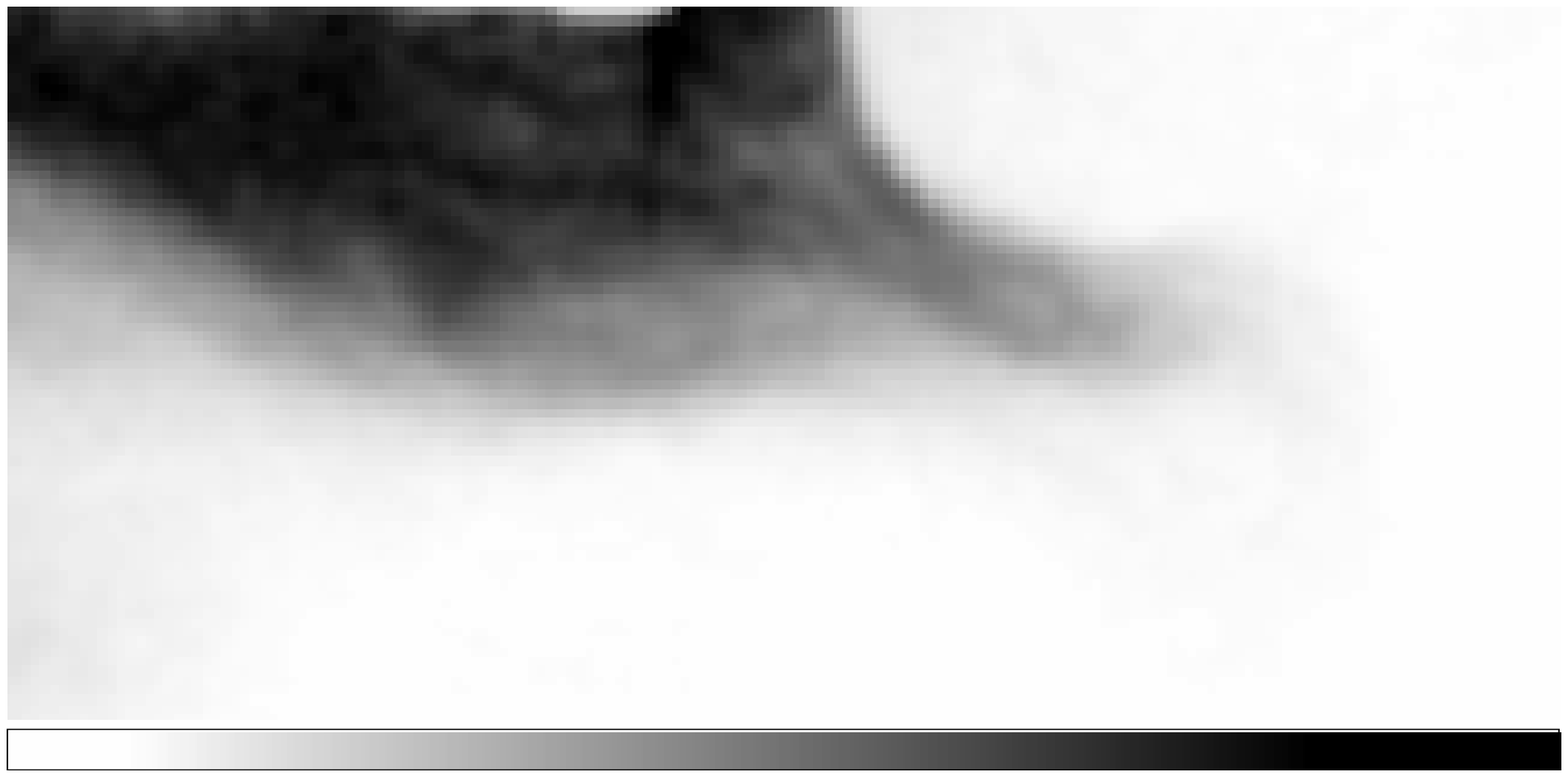}
\caption{The SW Finger showing regions used for spectral 
extraction. Regions are $5.9^{\prime\prime}
\times 11.8^{\prime\prime}$ 
or $10 \times 20$ light-weeks in size.  See caption of Figure 6 for
other detail.  Note that the photon index does not vary much along the length
of this feature.}
\end{figure}

\begin{figure}
\center
\includegraphics[totalheight=3in]{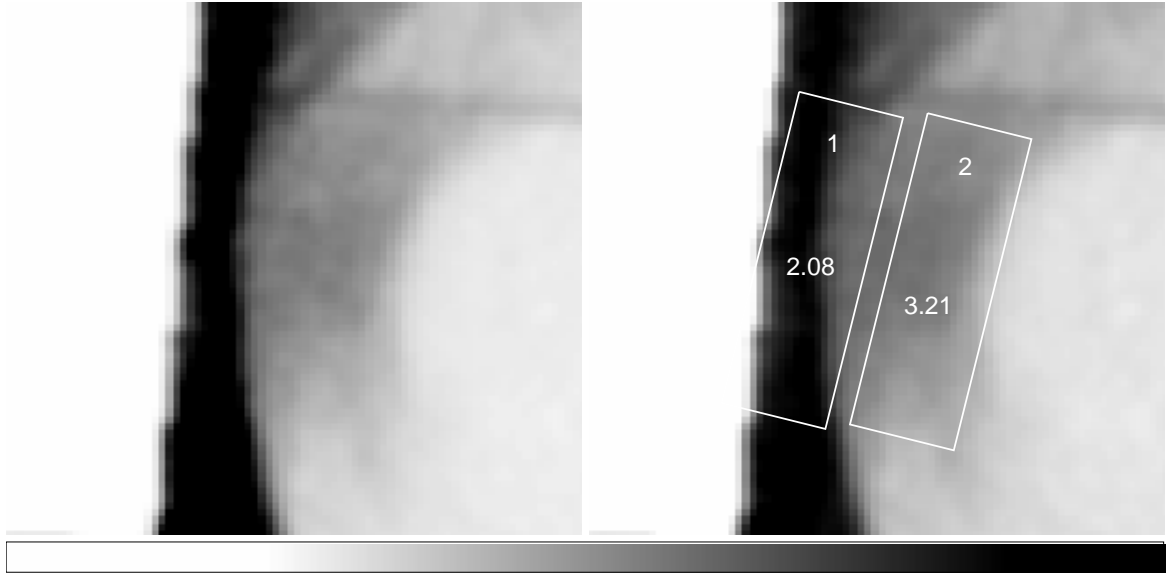}
\caption{The West Bay showing regions used for spectral 
extraction.  Regions are $7.4^{\prime\prime}
\times 22.1^{\prime\prime}$ in size.  The bright horizontal streak is the
CT streak from the Pulsar.} 
\end{figure}

\begin{figure}
\center
\includegraphics[totalheight=3in]{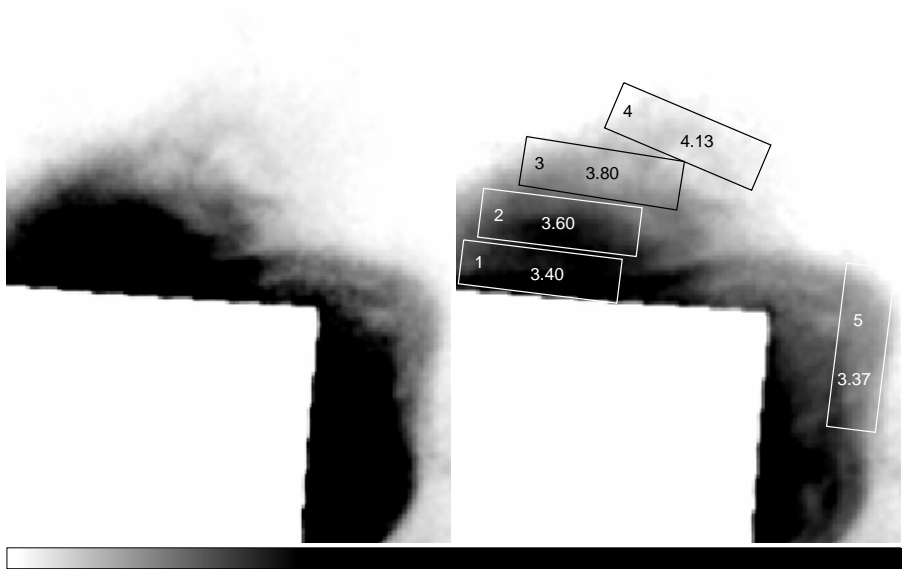}
\caption{The NW Loops area showing regions used for spectral 
extraction. Regions are $10^{\prime\prime}
\times 32^{\prime\prime}$ in size. } 
\end{figure}

\begin{figure}
\center
\includegraphics[totalheight=3in]{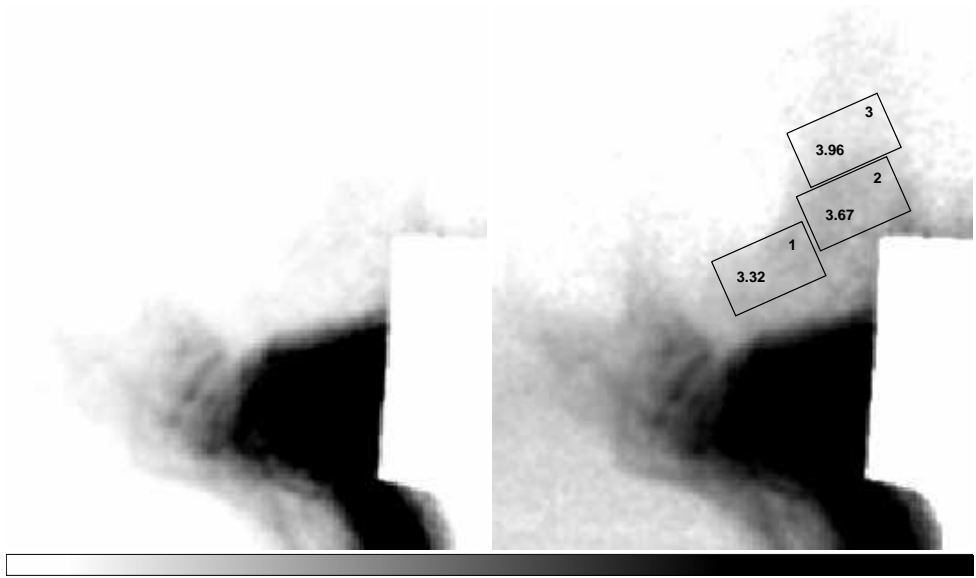}
\caption{The Chimney and NE Finger area showing regions used for spectral 
extraction.  Regions are $12^{\prime\prime}
\times 20^{\prime\prime}$ in size.  The faint diffuse emission filling
the image below the NW Finger is the CT streak from the bright central
part of the nebula. }
\end{figure}

\begin{figure}
\center
\includegraphics[totalheight=3in]{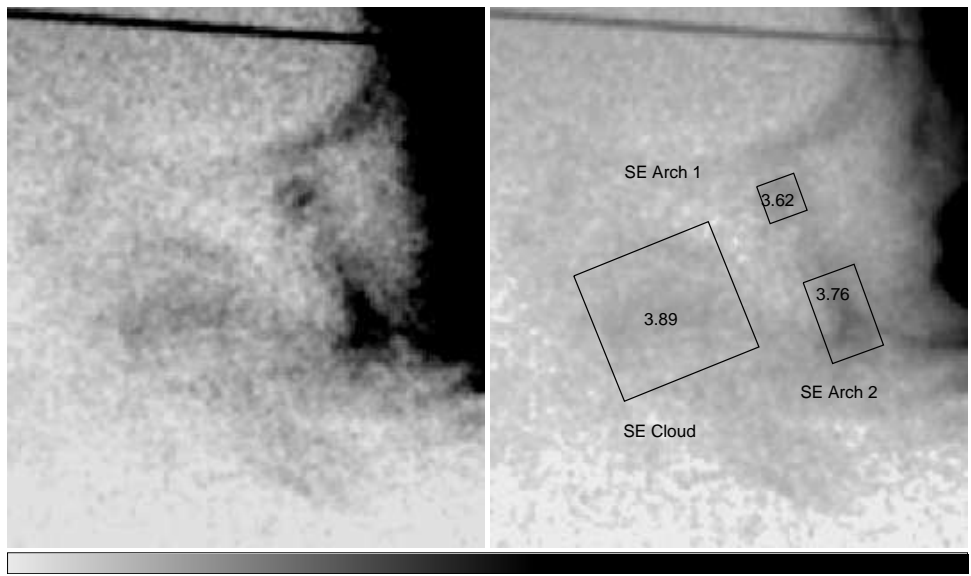}
\caption{The East Bay area showing regions used for spectral 
extraction.  Regions are $8^{\prime\prime}
\times 8^{\prime\prime}$, $11^{\prime\prime}
\times 17^{\prime\prime}$, and $27^{\prime\prime}
\times 29^{\prime\prime}$ in size.  The bright horizontal streak is the
CT streak from the Pulsar.} 
\end{figure}

\end{document}